 \documentclass[final,5p,times,twocolumn]{elsarticle}

\usepackage{amssymb}
\usepackage{lipsum}
\usepackage{amsmath}
\usepackage{hyperref}
\usepackage{xcolor}

\newcommand{\be}{\begin{equation}}
\newcommand{\ee}{\end{equation}}
\newcommand{\bea}{\begin{eqnarray}}
\newcommand{\eea}{\end{eqnarray}}
\newcommand{\mH}{\mathcal{H}}
\DeclareMathOperator{\Tr}{Tr}
\newcommand{\bS}{\mathbf{S}}
\newcommand{\mL}{\mathcal{L}}
\newcommand{\nn}{\nonumber}

\journal{Physics Letters B}

\begin{document}

\begin{frontmatter}



\title{State counting in gravity and maximal entropy principle}


\author[first]{Juan Hernandez}
\ead{hernanju@tcd.ie}
\affiliation[first]{organization={School of Mathematics and Hamilton Mathematics Institute, Trinity College},
            addressline={College Green}, 
            city={ Dublin 2},
            postcode={D02 PN40}, 
            country={Ireland}}

\author[second]{Mikhail Khramtsov}
\ead{mikhail.khramtsov311@gmail.com}
\affiliation[second]{organization={Department of Mathematical Physics, Steklov Mathematical Institute of Russian Academy of Sciences},
            addressline={ Gubkin str. 8}, 
            city={Moskow},
            postcode={119991},,
            country={Rusia}}

\begin{abstract}
It is known that the semiclassical approximation to the gravity path integral can be leveraged to explain certain inherently quantum aspects of gravity. One such aspect is the state-counting interpretation of the Bekenstein-Hawking entropy of black holes. A second aspect is the Page curve for the entanglement entropy of Hawking radiation, which agrees with expectations from unitarity. We show that these two questions are equivalent from the gravity path integral point of view. In particular, the Hawking's information loss puzzle gets resolved automatically by considering any (over)complete basis of black hole microstates which is compatible with black hole entropy. The tool which relates the two questions is a convex optimization problem for the von Neumann entropy of Hawking radiation.
\end{abstract}




\end{frontmatter}




\section{Introduction}

Black hole solutions in gravity are known to have a finite coarse-grained entropy given by the Bekenstein-Hawking formula $S_{BH} = \frac{A}{4G_N}$ \cite{Bekenstein:1973ur,Hawking:1975vcx}, which suggests that a black hole in quantum gravity should be treated as a quantum system with a Hilbert space of finite dimension $e^{S_{BH}}$. The Bekenstein-Hawking entropy appears already in the leading semiclassical order of the naively defined Euclidean General Relativity (GR) path integral \cite{Gibbons1977}.  While the state counting interpretation of the Bekenstein-Hawking entropy was explicitly confirmed in the UV-complete quantum gravity setting of string theory \cite{Strominger1996}, it has been a long-time puzzle to show that $S_{BH}$ has a state-counting interpretation in a low energy limit of a quantum theory of gravity in the semiclassical regime in more general settings. Recent work \cite{Balasubramanian:2022gmo,Balasubramanian:2022lnw,Climent2024,Balasubramanian:2024yxk,Balasubramanian:2025hns} leveraged the Euclidean path integral of GR to provide a statistical interpretation to the Bekenstein-Hawking entropy as the logarithmic dimension of the black hole Hilbert space spanned by an infinite family of semiclassical black hole microstates. 

Another famous puzzle in quantum gravity is the black hole information problem. In his seminal papers \cite{Hawking:1975vcx,Hawking:1976ra} Hawking argued that evaporation due to radiation of a black hole, that was formed from a pure state of matter, leads to a final state of radiation that is mixed. This violates unitarity of quantum evolution and creates information loss. This violation of unitarity would manifest as unbounded growth of the entanglement entropy of Hawking radiation collected by a distant observer, as the black hole evaporates. Page argued \cite{Page1993,Page19932} that if the total state of the system of black hole and its radiation is to remain pure, then the entanglement entropy of radiation has to be bounded, following what now is known as the Page curve. 
It is important to note that Hawking's information loss argument is based on consideration of the radiation in QFT on the black hole background, with the backreaction accounted for in a limited manner in the quasistatic approximation during black hole evaporation. The quantum degrees of freedom of the black hole itself, whose presence is suggested by the Bekenstein-Hawking entropy as mentioned above, are not accounted for in Hawking's information loss argument. 
Meanwhile, Page's entanglement entropy argument is based on representing the black hole and its radiation as a bipartite quantum mechanical system in a random entangled pure state, without explicitly including gravity. Reproducing the Page curve from gravity has been a long-standing problem, which has been resolved on a conceptual level recently by the replica wormhole mechanism \cite{Penington2022,Almheiri2020} (see \cite{Almheiri2021} for a review). In the replica wormhole computations, the late-time bounded phase of the Page curve emerges in the semiclassical Euclidean gravity path integral after accounting for nontrivial wormhole saddle points in the R\'enyi entropy computation. The internal black hole microstates only appear in the abstract form of EOW branes behind the horizon in the west coast model (PSSY) \cite{Penington2022}, and are not usually thought of as crucial in the gravitational Page curve computations.

    The aim of the present article is to show that the black hole entropy state counting puzzle and the Page curve puzzle are equivalent in general relativity at the nonperturbative level. Consequently, these puzzles are resolved at the semiclassical level by the same mechanism. In \cite{Balasubramanian:2022gmo} it was discussed that the overcompleteness of the family of semiclassical black hole microstates, constructed in that work, provides a unitary Page curve by replacing the family of EOW branes in the PSSY setup \cite{Penington2022}. This also essentially extends the PSSY analysis of the Page curve to GR in higher dimensions. In this article we make this argument precise by considering a maximization problem for the von Neumann entropy of radiation and using the state counting of the Bekenstein-Hawking entropy as one of the constraints for this maximization problem. We will explain that the unitary Page curve emerges as the unique solution to this maximization problem, for any family of black hole microstates which has the overcounting properties dictated by the Bekenstein-Hawking entropy. We will also argue that one can consider a reverse problem, which would maximize the black hole entropy, using the von Neumann entropy of radiation as a constraint. The solutions for both problems coincide. 

\section{State counting in gravity}
\label{sec: setup}

We begin the discussion by reviewing the black hole entropy state counting by the semiclassical black hole microstates introduced in~\cite{Balasubramanian:2022gmo}\footnote{These states are also higher dimensional generalizations of the partially entangled thermal states defined in JT gravity~\cite{Goel:2018ubv}.}. These semiclassical states are represented by black hole geometries with shells of matter behind the horizon. The geometries are constructed by gluing two solutions to the vacuum gravity equations along the shell worldvolumes. The vacuum gravity solutions, labeled here by $\alpha$, have metric in the form\footnote{We provide specific examples for the case of Einstein gravity for concreteness. More generally, higher derivative corrections would modify the blackening functions $f_\alpha(r)$, but we expect the qualitative picture remains the same.}
\begin{equation}
\begin{aligned}\label{eq: geometries}
ds^2_\alpha = f_\alpha(r) d\tau^2 + \frac{dr^2}{f_\alpha(r)} + r^2 d\Omega_{d-1}\,,\\ f_\alpha (r) =  \kappa r^2 + 1 - \frac{16 \pi G M_\alpha}{(d-1) V_\Omega r^{d-2}}\,,
\end{aligned}
\end{equation}
where $\kappa = 1$ corresponds to asymptotically AdS$_{d+1}$ and $\kappa=0$ to asymptotically flat geometries, see~\cite{Balasubramanian:2022lnw,Balasubramanian:2025zey}.
The ADM masses $M_\alpha$ determine whether the geometry on either side of the shell is a black hole ($M_\alpha>0$) or empty AdS/flat space ($M_\alpha=0$). Between the two regions is the worldvolume of a thin spherical shell of matter, with stress tensor $S_{ab} = \sigma u_a u_b$, where $\sigma$ is the fluid density and $u^a$ is the unit normalized fluid velocity. The mass of the shell is conserved along the worldvolume, and is given by $m= \sigma V_\Omega r^{d-1}$.

The trajectory of the shell, and therefore of the interface between the two geometries~\eqref{eq: geometries}, can be found via the Israel junction conditions~\cite{Israel:1966rt}, which amount to getting the same induced metric from the two geometries and balancing the difference in extrinsic curvatures with the stress energy of the shell of matter (see \cite{Balasubramanian:2024rek} for details). 
For spherically symmetric configurations, the we can use the embedding $\tau= \tau_\alpha(T)$ and $r=R(T)$, where $T$ is the proper synchronous time of the shell and the junction conditions reduce to a conservation equation for a particle in a potential $\dot{R}^2 +V_{\rm eff}(r) = 0$, where $V_{\rm eff}$ is an effective potential derived from the geometries~\eqref{eq: geometries}, see~\cite{Balasubramanian:2022gmo,Balasubramanian:2022lnw,Balasubramanian:2024rek}.

The states constructed as explained above have the same geometry between the horizon and the asymptotic boundary, and are indistinguishable from a standard black hole geometry for an observer outside of the horizon. For this reason, they are microstates of the corresponding black hole Hilbert space. Note that the external black hole geometry is independent of the mass of the shell that separates the two regions, and one can therefore build an infinite number of such semiclassical black hole microstates for a given black hole.
Moreover, one can estimate their overlap using the gravity path integral~\cite{Balasubramanian:2022gmo}, and at leading order in the semiclassical approximation, they are orthogonal as long as the masses of the shells are macroscopically different. For example, we can make a family of semiclassical black hole microstates $\left\{|\Psi_i\rangle\right\}_{i=1}^\Omega$, where each state is built as explained above but with different shell masses, given by $m_i = i \mu$ for some $\mu$ that is much larger than the Planck mass. 

A central object to understand the properties of the Hilbert space spanned by such a family of states is the Gram matrix $G_{ij} = \langle \Psi_i | \Psi_j \rangle$.
The semiclassical approximation to the gravity path integral can be leveraged to find, after appropriate normalization of the states~\cite{Balasubramanian:2022gmo,Balasubramanian:2022lnw}
\begin{equation}\label{eq: moments}
    \overline{G_{ij}} = \delta_{ij}\,, \quad \left. \overline{G_{ik_1}G_{k_1 k_2} \cdots G_{k_{n-1}j}}\right|_{\rm con.} = \delta_{ij}  \frac{Z(n\beta)^\nu}{Z(\beta)^{\nu n}}\,,
\end{equation}
where $\nu=1$ if the geometry on the other side of the shell is the vacuum solution ($M_\alpha=0$), and $\nu=2$ otherwise. The second equation in~\eqref{eq: moments} denotes the connected contributions to the moments of overlaps, for example $\left.\overline{G_{ik}G_{kj}}\right|_{\rm con.}  = \overline{G_{ik} G_{kj}} -  \overline{G_{ik}} \,\overline{G_{kj}}$. The fact that the connected part of the higher moments of overlaps are non-zero directly imply that the basis of semiclassical black hole microstates are in fact non-orthogonal.

A useful tool is to work with microcanonical states $|\Psi_i^E\rangle$ by projecting $|\Psi_i\rangle$ into a microcanonical window, which leads to replacing canonical expressions like $Z(n\beta)$ with their inverse Laplace transforms (see \cite{Balasubramanian:2024yxk} for details).
The resulting semiclassical approximations for the moments of overlaps are
\begin{equation}\label{eq: micro moments}
     \overline{G^E_{ij}} = \delta_{ij}\,, \quad \left. \overline{G^E_{ik_1}G^E_{k_1 k_2} \cdots G^ E_{k_{n-1}j}}\right|_{\rm con.}= \delta_{ij} e^{\nu(1-n) \mathbf{S}(E)}\,,
\end{equation}
where $\mathbf{S}(E)$ is the microcanonical entropy. The second equation in~\eqref{eq: micro moments} implies that there is a non-perturbatively small overlap between different states. This non-perturbative overlap can cause the dimension of the Hilbert space $\mH_{BH} = \text{Span}\left\{ |\Psi^E_i\rangle \right\}_{i=1}^\Omega$ to be less than the number of basis elements $\Omega$. This can be neatly captured by the resolvent method.

The resolvent of a given matrix $G$ is
\begin{equation}
\label{eq: resolvent def}
    R_{ij}(\lambda) = \left(\frac{1}{\lambda-G}\right)_{ij} = \frac{\delta_{ij}}{\lambda} + \frac{1}{\lambda} \sum_{n=1}^\infty \frac{G^n_{ij}}{\lambda^n}\,.
\end{equation}
The poles of the resolvent trace $R(\lambda)$ are located at the eigenvalues of $G_{ij}$, and their residues count the number of eigenstates with the corresponding eigenvalues. To find the dimension of the Hilbert space spanned by the semiclassical microstates, we need to find the rank of the Gram matrix. 
This is equivalent to finding the residue at the origin of the resolvent trace, since it counts the dimension of the null space of $G_{ij}$.  The dimension of the spanned Hilbert space is ${\rm dim} \left({\cal H}_{BH}\right)  = \Omega - {\rm Res}\left( R,0\right) $.

From the second equality in~\eqref{eq: resolvent def} and naively using the first equation in~\eqref{eq: micro moments} to approximate the Gram matrix, omitting the non-vanishing connected contributions that are found using the gravity path integral, one finds 
\begin{equation}
    R(\lambda) = \frac{\Omega}{1-\lambda}  \Rightarrow {\rm Res}\left( R,0\right)=0\,,
\end{equation}
which leads to the dimension of the spanned Hilbert space being $\Omega$. However, as more and more states are added to the basis by increasing $\Omega$, eventually the small overlaps between the states affect the dimension of the spanned Hilbert space. Indeed, for very large $\Omega$, the connected contributions are in fact dominant over the disconnected contributions to the moments of the Gram matrix. In this limit, the resolvent trace~\eqref{eq: resolvent def} is instead given by
\begin{equation}\label{eq: R trace}
    R(\lambda) \approx \frac{\Omega}{\lambda} \left( 1+ \frac{1}{\lambda - \Omega e^{- \nu \mathbf{S}}}\right)  \Rightarrow  {\rm Res}\left( R,0\right) = \Omega-e^{\nu \mathbf{S}}\,,
\end{equation}
and the Hilbert space dimension saturates at $e^{\nu \mathbf{S}}$, in agreement with the statistical interpretation of the Bekenstein-Hawking entropy of the black hole. The value $\nu=1,2$ depends on whether we are spanning the Hilbert space of a single or a double-sided black hole. Note that the only other pole of the resolvent trace~\eqref{eq: R trace} is at $\Omega e^{-\nu \mathbf{S}}$, and the residue is $e^{\nu \mathbf{S}}$. We will see in section \ref{sec: entropy} that the corresponding distribution of eigenvalues is precisely the one that maximizes the von Neummann entropy in a simple radiation model if one imposes the residue at the origin as in the second equation in~\eqref{eq: R trace}. The similarity between the eigenvalue distribution derived from~\eqref{eq: R trace} and the entropy maximization argument in section~\ref{sec: entropy} is due to including only the fully connected contributions to the moments of the Gram matrix. This result corresponds to the $\Omega e^{-\nu \mathbf{S}}\to \infty$ limit of the more general Marchenko-Pastur distribution found in~\cite{Balasubramanian:2022gmo} by including all contributions. 


\section{Maximizing the entropy of Hawking radiation}\label{sec: entropy}

To include the Hawking radiation in our consideration, we treat the black hole and its radiation as a bipartite quantum system with Hilbert space $\mH_{BH} \otimes \mH_R$. Following \cite{Penington2022,Balasubramanian:2022gmo}, we consider the state
\be
| \Phi \rangle = \frac{1}{\sqrt{\Omega}} \sum_{i=1}^\Omega |\Psi_i \rangle \otimes |i \rangle_R\,. \label{eq: total state}
\ee
The states $|\Psi_i \rangle$ are the black hole microstates constructed as explained in section~\ref{sec: setup}, and $\mH_{BH} = \text{Span}\left\{ |\Psi^E_i\rangle \right\}_{i=1}^\Omega$. The states $|i\rangle_R$ are an orthonormal basis of the Hawking radiation degrees of freedom. We write the state (\ref{eq: total state}) in the naively maximally entangled form with the goal to find the upper bound for the von Neumann entropy of radiation
\be
S_R = -\Tr \rho_R \log \rho_R\,, \label{eq: von Neumann entropy}
\ee
where $\rho_R = \Tr_{BH} | \Phi \rangle \langle \Phi |$ is the radiation density matrix with the components given by 
\be
(\rho_R)_{ij} = \frac{\langle \Psi_i | \Psi_j \rangle}{\Omega} \equiv \frac{G_{ij}}{\Omega} \,.
\ee
For the black hole microstates $|\Psi_i \rangle$ described in section \ref{sec: setup}, the state (\ref{eq: total state}) describes a black hole which is in equilibrium with an external non-gravitational bath which collects the Hawking radiation. 
The number of radiation states $\Omega$ can serve as a time parameter, under the assumption of quasi-static black hole evaporation. In the present article we focus on the eternal black hole case, and we will consider different values of $\Omega$ compared to $e^{S_{BH}}$. 

To study the von Neumann entropy of the radiation, it is convenient to introduce the density of Gram matrix eigenvalues as the discontinuity of the resolvent across the real axis:
\be
D(\lambda) = \lim_{\epsilon \to 0} \frac{1}{2\pi i} (R(\lambda - i\epsilon) - R(\lambda + i\epsilon))\,.
\ee
Since the resolvent is by definition (\ref{eq: resolvent def}) a meromorphic function with all poles on the real non-negative axis, then all closed contour integrals of the resolvent (times any holomorphic function) can be written as integrals over the real positive half-axis with the resolvent replaced by $D(\lambda)$\footnote{The integral over $D(\lambda)$ is nontrivial only for $\lambda > 0$ because all singularities of the resolvent are on the non-negative half-axis.}. To this end, the number of poles of the resolvent can be counted by the integral
\begin{equation}\label{eq: C1}
    \textbf{C1}:\quad \oint \frac{d\lambda}{2\pi i}\ R(\lambda) = \int_0^{+\infty} d\lambda\  D(\lambda) = \Omega\,, \tag{C1}
\end{equation}
where the integral of the resolvent is taken over the closed contour which encompasses all of the singularities of the resolvent. Thus we have a contour integral constraint for $R(\lambda)$ or, equivalently, a real integral constraint for $D(\lambda)$. Recall that we also have $\Tr G = \Omega$, which gives another integral constraint
\begin{equation}\label{eq: C2}
    \textbf{C2}:\quad \oint \frac{d\lambda}{2\pi i} \lambda R(\lambda)= \int_0^{+\infty} d\lambda\  \lambda D(\lambda)  =\Omega, \tag{C2}
\end{equation}
for the same contour as in~\eqref{eq: C1}. As a final constraint, we require that the recall that the eigenvalue density has to be non-negative
\begin{equation}
    \textbf{C3}: \quad D(\lambda) \geq 0\,. \tag{C3} \label{eq: C3}
\end{equation}

The von Neumann entropy can be written in terms of the density of eigenvalues as\footnote{One can derive this formula by using the standard replica trick $S =  \lim_{n \to 1} \frac{\log \Tr \rho^n}{1-n}$ and using the contour integral of the resolvent as $\Tr \rho^n = \oint \frac{d\lambda}{2\pi i} \left(\frac{\lambda}{\Omega}\right)^n R(\lambda)$.}
\be
S = - \int_0^{+\infty} d\lambda\ \frac{\lambda}{\Omega} \log \frac{\lambda}{\Omega} D(\lambda)\,. \label{eq: S - D}
\ee
The goal is to maximize the functional $S$ over $D(\lambda)$, where $D(\lambda)$ is subject to constraints. This is an infinite-dimensional linear optimization problem with equality constraints and one inequality constraint \cite{boyd2004convex}. We explain the solution of this maximization problem in the appendix \ref{sec: maximization SvN appendix}. Here we will now explain the results in two cases depending on the value of $\Omega$.

\textbf{Case 1. $\Omega < e^\bS$}. In this case the microstate basis $\{ |\Psi_i \rangle\}$ is complete and the resolvent has no pole at the origin. We have to maximize the von Neumann entropy functional (\ref{eq: S - D}) subject to the constraints \ref{eq: C1}-\ref{eq: C3}. 

Thus the maximizing eigenvalue density corresponds to a standard maximally mixed state with $D_{\text{Hawking}}(\lambda) = \Omega \delta(\lambda - 1)$, and the maximal von Neumann entropy is given by 
\be
S_{\text{Hawking}} = - \int_0^{+\infty} d\lambda\ \frac{\lambda}{\Omega} \log \frac{\lambda}{\Omega} D_{\text{Hawking}}(\lambda) = \log \Omega\,. \label{eq: Shawking}
\ee
This matches the early-time result for Page curve \cite{Page1993}, often called in the literature the Hawking phase.

\textbf{Case 2. $\Omega > e^\bS$}. In this case the microstate basis $\{ |\Psi_i \rangle\}$ is overcomplete, and the resolvent gains a pole at the origin as shown in (\ref{eq: R trace}). The constraints \ref{eq: C1}-\ref{eq: C3} remain, however now we have to add an additional equality constraint: 
\be
\textbf{C4}: \quad \oint \frac{d\lambda}{2\pi i} R(\lambda) = \int_{0-\epsilon}^{0+\epsilon} d\lambda D(\lambda) = \Omega - e^{\nu \bS}\,, \label{eq: C4}
\tag{C4}
\ee
where the contour integral is taken over an infinitesimal contour around the origin. 

The extremizing eigenvalue density then reads 
\be
D_{\text{Page}} = (\Omega - e^{\nu \bS}) \delta(\lambda) + e^{\nu \bS} \delta(\lambda - \Omega e^{-\nu \bS})\,. 
\ee
The resulting von Neumann entropy is
\be
S_{\text{Page}} =  - \int_0^{+\infty} d\lambda\ \frac{\lambda}{\Omega} \log \frac{\lambda}{\Omega} D_{\text{Page}}(\lambda) = \nu \bS\,. \label{eq: Spage}
\ee
This is exactly the late time bound of the Page curve for an eternal black hole \cite{Page19932,Almheiri2021}. 

Note that the Page curve we obtained from the maximization problem has a sharp transition, because we have been working in the microcanonical ensemble. In the canonical ensemble the Page curve is not tractable analytically in general theory of gravity, but the computationally accessible examples show that the Page transition gets smoothed in the canonical ensemble \cite{Penington2022,Balasubramanian:2022gmo}. 

\section{Discussion}

We have shown that the unitary Page curve for an eternal black hole in an asymptotically flat or AdS spacetime of any dimension arises from solving the linear optimization problem with respect to eigenvalue density of the microstate Gram matrix. This problem is defined on a set of functions, which is defined by the microstate basis overcounting properties in the microcanonical ensemble, as well as by natural positivity and Gram matrix trace constraints. 

One can also see that it is possible to reverse the argument to obtain the dimension of the Hilbert space, which is spanned by the black hole microstates $|\psi_k \rangle$, from a maximization condition using the von Neumann entropy obeying the Page curve as a constraint. To do this, one can consider the functional $\Sigma = \int_{0+\epsilon}^{+\infty} d\lambda\ D(\lambda)$, which counts the number of non-zero eigenvalues of the Gram matrix, i. e. the dimension of the black hole Hilbert space. Then one can maximize $\Sigma$, using the von Neumann entropies (\ref{eq: Shawking}) and (\ref{eq: Spage}) as the constraints. We give the details in appendix \ref{sec: Sigma appendix}. As the result, for $\Omega < e^{\nu \bS}$ we get $\Sigma = \Omega$, and for $\Omega > e^{\nu \bS}$ we get $\Sigma= e^{\nu \bS}$. This gives the dimension of the black hole Hilbert space using the von Neumann entropy of radiation as an input. The von Neumann entropy maximization problem from section \ref{sec: entropy} together with this reverse optimization problem having the same solution for $D(\lambda)$ shows that the black hole entropy state counting and radiation entanglement entropy are equivalent problems.

The optimization problem for the von Neumann entropy is universal in the sense that it does not depend on the specific family of black hole microstates. The family of semiclassical shell microstates, introduced in \cite{Balasubramanian:2022gmo} and reviewed in section \ref{sec: setup}, leverages the semiclassical limit of the gravity path integral in order to derive the resolvent residue at the origin (\ref{eq: R trace}), which in turn determines the dimension of the null subspace in the span of the family of microstates. However, the semiclassical limit is not needed when going from the resolvent residue to the Page curve, and so our optimization argument is nonperturbative and will hold for any family of states prepared by the Euclidean gravity path integral whose dimension saturates at the value of $e^{\nu \bS}$ due to their overlap structure. The entropy maximization argument is also an improvement over the Page's argument \cite{Page1993}, because we do not require an orthonormal basis and a Schmidt decomposition for the full state $|\Phi\rangle$ (\ref{eq: total state}). 

The Page curve derived from the entropy maximization does not exactly match the Page curve of the gravitational models \cite{Penington2022}, but it bounds it from above, and in fact coincides with the coarse grained entropy curve computed in \cite{Balasubramanian:2022gmo}. As mentioned in the end of section \ref{sec: setup}, the density of states derived in those gravitational models matches the solution to the maximization problem in the $\Omega e^{-\nu \mathbf{S}} \to \infty$ limit. The von Neumann entropy is sensitive to the structure of the density of states throughout its whole spectrum, while the coarse-grained entropy is only sensitive to its residue at the origin.

Our argument has two main limitations. The first limitation is that in the model described by the state (\ref{eq: total state}) the observer does not appear as part of the gravitational spacetime setup. The second limitation is that this model does not describe real-time dynamics of black hole evaporation, but instead approximates a state of the system during quasi-static evolution. It would be interesting to consider the entropy maximization argument in models with dynamics and physical observers.

\section*{Acknowledgements}

The authors are grateful to Vijay Balasubramanian, Ben Craps, Dongming He, Maria Knysh and Tom Yildirim for fruitful collaborations and discussions on semiclassical black hole microstates. M. K. is also thankful to  I. Aref'eva and I. Volovich for useful discussions on mathematical foundations.  The work of JH is supported by Taighde Eireann – Research Ireland under Grant number SFI-22/FFP-P/11444. The work of M. K. was supported by the Russian Science Foundation under grant no. 24-11-00039, https://rscf.ru/en/project/24-11-00039/, and performed at Steklov Mathematical Institute of Russian Academy of Sciences.

\appendix
\section{Solution of the optimization problems}

\subsection{Maximizing the von Neumann entropy}
\label{sec: maximization SvN appendix}

\textbf{Case 1. $\Omega < e^\bS$}. In this case the microstate basis $\{ |\Psi_i \rangle\}$ is complete and the resolvent has no pole at the origin. We have to maximize the von Neumann entropy functional (\ref{eq: S - D}) subject to the constraints \ref{eq: C1}-\ref{eq: C3}. To solve this maximization problem, we introduce the Lagrangian
\be
\begin{aligned}
\label{eq: L}
\mL & = \alpha \left[\int_0^{+\infty} d\lambda D(\lambda) - \Omega\right] + \beta \left[\int_0^{+\infty} d\lambda \lambda D(\lambda) - \Omega\right] \\  & \quad \quad \quad \quad + S + \int_0^{+\infty} d\lambda \Gamma(\lambda) D(\lambda)\,,
\end{aligned}
\ee
where $S$ is defined in (\ref{eq: S - D}). Here $\alpha$, $\beta$ and $\Gamma(\lambda)$ are Lagrange multipliers which enforce the constraints \ref{eq: C1}-\ref{eq: C3}. Varying with respect to $D(\lambda)$ gives the extremality condition
\be
-\frac{\lambda}{\Omega} \log \frac{\lambda}{\Omega} + \alpha + \beta \lambda + \Gamma(\lambda) = 0\,. \label{eq: extemality S}
\ee
Note that this equation can only be satisfied for all $\lambda$ with nontrivial $\Gamma(\lambda)$ -- in other words, the optimization problem has no solution without the positivity constraint. Furthermore, the extremal $D(\lambda)$ are non-zero only for the values of $\lambda$ for which $\Gamma(\lambda)=0$.

The positivity condition \ref{eq: C3} implies that $\Gamma(\lambda) \geq 0$. 
Furthermore, the function $\Gamma(\lambda)$ is convex for $\lambda >0$, and together with the above inequality constraint, this means that  $\Gamma(\lambda)$ can have at most a single zero which satisfies $\Gamma(\lambda_*) = \Gamma'(\lambda_*) = 0$. In other words, the function $\Gamma(\lambda)$ has to be non-negative and touch the zero axis from above at its extremal point. We have the extremality condition for $\Gamma$:
\be
\Gamma'(\lambda_*) = \frac{1}{\Omega} \log \frac{\lambda_*}{\Omega} + \frac{1}{\Omega} - \beta = 0 \quad \Rightarrow \quad \lambda_* = \Omega e^{\beta \Omega -1}\,. \label{eq: f'}
\ee
The extremal value of $\Gamma$ has to be zero: 
\be
\Gamma(\lambda_*) =- \alpha - e^{\beta \Omega -1} = 0 \quad \Rightarrow \quad \alpha = -e^{\beta \Omega -1}\,. \label{eq: f}
\ee

We have seen that the extremality condition (\ref{eq: extemality S}) can only be satisfied with $\Gamma(\lambda)=0$ at a single point. This means that the solution for $D(\lambda)$ has to be a delta function: 
\be
D(\lambda) = A \delta(\lambda - \lambda_*)\,.
\ee
From the constraints \ref{eq: C1} and \ref{eq: C2} it immediately follows that $\lambda_* = 1$ and $A = \Omega$. Using (\ref{eq: f}) and (\ref{eq: f'}), we determine the Lagrange parameters:
\be
\beta = \frac{1}{\Omega} + \frac{1}{\Omega} \log \frac{1}{\Omega}\,, \qquad \alpha = -\frac{1}{\Omega}\,. \label{eq: alphabeta}
\ee
The result for the extremal density of eigenvalues is 
\be
D_{\text{Hawking}}(\lambda) = \Omega \delta(\lambda - 1)\,. \label{eq: Dhawking}
\ee
Substituting this and with the $\alpha$, $\beta$ given by (\ref{eq: alphabeta}) into the Lagrangian (\ref{eq: L}), we find the maximal von Neumann entropy:
\be
S_{\text{Hawking}} = - \int_0^{+\infty} d\lambda\ \frac{\lambda}{\Omega} \log \frac{\lambda}{\Omega} D_{\text{Hawking}}(\lambda) = \log \Omega\,. \label{eq: Shawking 2}
\ee

\textbf{Case 2. $\Omega > e^\bS$}. The constraints \ref{eq: C1}-\ref{eq: C3} in this case get supplemented by the extra equality constraint \ref{eq: C4} which requires $R(\lambda)$ to have a pole at the origin. The Lagrangian now takes the form
\be
\begin{aligned}
\label{eq: L 2}
 &\mL =  \alpha \left[\int_0^{+\infty} d\lambda D(\lambda) - \Omega\right] + \beta \left[\int_0^{+\infty} d\lambda \lambda D(\lambda) - \Omega\right] \\
 & + S +  \chi \left[\int_{-\epsilon}^{+\epsilon} d\lambda D(\lambda) - \Omega + e^{\nu \bS}\right] + \int_0^{+\infty} d\lambda \Gamma(\lambda) D(\lambda)\,,
\end{aligned}
\ee
Note that the integration in the $\chi$-term is supported only on an infinitesimal interval around zero. Therefore, in this case the extremality condition $\frac{\delta \mL}{\delta D(\lambda)}=0$ reads
\bea
-\frac{\lambda}{\Omega} \log \frac{\lambda}{\Omega} + \alpha + \beta \lambda +  \Gamma(\lambda) &=& 0\,, \quad \lambda > 0\,,\nn\\
\alpha + \chi + \Gamma(0) &=& 0\,.
\label{eq: extremality Page}
\eea
The constraint \ref{eq: C4} on its own implies that a feasible $D(\lambda)$ gains a delta function at the origin. Indeed, in addition to $\Gamma(\lambda_*) = 0$ as before, the second equation in (\ref{eq: extremality Page}) allows us to set $\Gamma(0) = 0$ by choosing $\chi =-\alpha$. So the solution has a general form
\be
D(\lambda) = B \delta(\lambda) + A \delta(\lambda - \lambda_*)\,.
\ee
The constraint \ref{eq: C4} fixes $B = \Omega - e^{\nu \bS}$. To determine the location of $\lambda_*$, we can subtract \ref{eq: C4} from \ref{eq: C1} to get
\be
\int_{+\epsilon}^{+\infty} d\lambda D(\lambda) = e^{\nu \bS}\,.
\ee
Together with \ref{eq: C2}, this fixes $\lambda_* = \Omega e^{-\nu \bS}$ and $A = e^{\nu \bS}$. Thus the extremizing eigenvalue density then reads 
\be \label{eq: Dpage}
D_{\text{Page}} = (\Omega - e^{\nu \bS}) \delta(\lambda) + e^{\nu \bS} \delta(\lambda - \Omega e^{-\nu \bS})\,. 
\ee
The values of the parameters $\alpha$ and $\beta$ are determined from the r.h.s. of the equations (\ref{eq: f'}),(\ref{eq: f}) using $\lambda_* = \Omega e^{-\nu \bS}$. The Lagrangian on this solution equals the maximal von Neumann entropy, which is
\be
S_{\text{Page}} =  - \int_0^{+\infty} d\lambda\ \frac{\lambda}{\Omega} \log \frac{\lambda}{\Omega} D_{\text{Page}}(\lambda) = \nu \bS\,. \label{eq: Spage 2}
\ee

\subsection{Maximizing the dimension of black hole Hilbert space}
\label{sec: Sigma appendix}

Here we consider the problem to maximize the dimension of the black hole Hilbert space $\mH_{BH} = \text{Span}\left\{ |\Psi^E_i\rangle \right\}_{i=1}^\Omega$ using the radiation Page curve as a constraint. Specifically, we aim to maximize the functional 
\be
\Sigma = \int_{0+\epsilon}^{+\infty} d\lambda\ D(\lambda)
\ee
subject to constraints \ref{eq: C1}-\ref{eq: C3} as well as the von Neumann entropy constraint
\be
- \int_0^{+\infty} d\lambda\ \frac{\lambda}{\Omega} \log \frac{\lambda}{\Omega} D(\lambda) = S_0\,.
\ee
The Lagrangian for the maximization problem reads 
\be
\begin{aligned}
\mL &=  \Sigma + \theta (S - S_0) + \int_0^{+\infty} d\lambda \Gamma(\lambda) D(\lambda) \\ &  +\alpha \left[\int_0^{+\infty} d\lambda D(\lambda) - \Omega\right] + \beta \left[\int_0^{+\infty} d\lambda \lambda D(\lambda) - \Omega\right]\,.
\end{aligned}
\ee
The extremality condition is obtained by varying over $D(\lambda)$:
\bea
1 -\theta \frac{\lambda}{\Omega} \log \frac{\lambda}{\Omega} +\alpha + \beta \lambda + \Gamma(\lambda) &=& 0\,, \quad \lambda > 0\,,\nn\\
\alpha + \Gamma(0) &=& 0\,. \label{eq: extremality Sigma}
\eea
By the same arguments as above, the extremality conditions (\ref{eq: extremality Sigma}) together with the inequality constraint $\Gamma(\lambda) \geq 0$ imply that the solution for $D(\lambda)$ has to have the form 
\be
D(\lambda) = B \delta(\lambda) + A \delta(\lambda - \lambda_*) \label{eq: D ansatz}
\ee
with $\lambda_* > 0$. Then the condition $\Gamma(0)=0$ sets $\alpha=0$ immediately via the second equation in (\ref{eq: extremality Sigma}). As above in (\ref{eq: f'})-(\ref{eq: f}), we have the equations
\be
\Gamma(\lambda_*) = \Gamma'(\lambda_*)=0\,.
\ee
These two equations can be solved together with the von Neumann entropy constraint $S=S_0$ evaluated on the ansatz (\ref{eq: D ansatz}) to determine $\lambda_*$, $\beta$, $\theta$ and $A$:
\bea
&&\lambda_* = \Omega e^{-S_0}\,,\qquad \theta = -e^{S_0}\,,\\
&&\beta = \frac{1}{\Omega} (S_0-1)e^{S_0}\,, \qquad A = e^{S_0}\,.
\eea

The constraint \ref{eq: C1} yields
\be
B + A = \Omega \quad \Rightarrow B = \Omega - e^{S_0}\,.
\ee
We now proceed to determine $A$, $B$ and $\lambda_*$ for two cases depending on the value of the constant $S_0$. 

\textbf{Case 1. $S_0 = \log \Omega$}. In this case we find the parameteres of the solution
\be
\lambda_* = 1\,, \qquad A = \Omega\,, \qquad B = 0\,.
\ee
We find that the solution is given by $D_{\text{Hawking}}$ in (\ref{eq: Dhawking}), and we find 
\be
\Sigma_{\text{Hawking}} = \Omega\,.
\ee

\textbf{Case 2. $S_0 = \nu\mathbf{S} < \log \Omega$}. In this case we find
\be
\lambda_* = \Omega e^{-\nu \bS} \,, \qquad A = e^{\nu \bS}\,, \qquad B = (\Omega - e^{\bS})\,.
\ee
We find that the solution is given by $D_{\text{Page}}$ in (\ref{eq: Dpage}), and we find 
\be
\Sigma_{\text{Page}} = e^{\nu\mathbf{S}} \,.
\ee

\bibliographystyle{elsarticle-harv} 
\bibliography{essay_refs}






\end{document}